\documentclass[aps,preprint,showpacs,preprintnumbers,eqsecnum,amsmath,amssymb]{revtex4}

\usepackage{graphics,epsfig}
\usepackage{graphicx}
\usepackage{dcolumn}
\usepackage{bm}

\begin{document}

\title{Quasinormal modes of Dirac field perturbation
in Schwarzschild-anti-de Sitter black hole}
\author{Jiliang Jing} \email{jljing@hunnu.edu.cn}
\affiliation{ Institute of Physics and  Department of Physics, \\
Hunan Normal University,\\ Changsha, Hunan 410081, P. R. China }

\baselineskip=0.65 cm

\vspace*{0.2cm}
\begin{abstract}
\vspace*{0.2cm}

The quasinormal modes (QNMs) associated with the decay of Dirac
field perturbation around a Schwarzschild-anti-de Sitter (SAdS)
black hole is investigated by using Horowitz-Hubeny approach. We
find that both the real and the imaginary parts of the fundamental
quasinormal frequencies for large black holes are linear functions
of the Hawking temperature, and the real part of the fundamental
quasinormal frequencies for intermediate and small black holes
approximates a temperature curve and the corresponding imaginary
part is almost a linear function of the black hole radius. The
angular quantum number has the surprising effect of increasing the
damping time scale and decreasing the oscillation time scale for
intermediate and small black holes. We also find that the rates
between quasinormal frequencies and black hole radii for large,
intermediate and small black holes are linear functions of the
overtone number, i. e., all the modes are evenly spaced.

\end{abstract}

 \vspace*{1.5cm}

\maketitle

\section{introduction}

It is well known that the quasinormal modes (QNMs) of a black hole
is defined as proper solutions of the perturbation equations
belonging to certain complex characteristic frequencies which
satisfy the boundary conditions appropriate for purely ingoing
waves at the event horizon and purely outgoing waves at infinity
\cite{Chand75}. The original interest in QNMs of black hole arose
due to they are characteristics of black holes which do not depend
on initial perturbations. It is generally believed that QNMs carry
a unique footprint to directly identify the existence of a black
hole. Through the QNMs, one can extract information about the
physical parameters of the black hole---mass, electric charge, and
angular momentum---from the gravitational wave signal by fitting
the few lowest observed quasinormal frequencies to those
calculated from the perturbation analysis, as well as test the
stability of the event horizon against small perturbations.
Recently, this interest has been renewed since the QNMs may be
related to fundamental physics, such as the thermodynamic
properties of black holes in loop quantum gravity \cite{Hod}
\cite{Dreyer}  since the real part of quasinormal frequencies with
a large imaginary part for the scalar field in the Schwarzschild
black hole is equal to the Barbero-Immirzi parameter
\cite{Hod}\cite{Dreyer}\cite{Baez}\cite{Kunstatter}, a factor
introduced by hand in order that loop quantum gravity reproduces
correctly the entropy of the black hole, and the QNMs of anti-de
Sitter (AdS) black holes have a direction interpretation in terms
of the dual conformal field theory (CFT) \cite{Maldacena}
\cite{Witten} \cite{Kalyana}. According to the AdS/CFT
correspondence, a large static black hole in asymptotically AdS
spacetime corresponds to a thermal state in CFT, and the decay of
the test field in the black hole spacetime corresponds to the
decay of the perturbed state in CFT. The dynamical timescale for
the return to thermal equilibrium can be done in AdS spacetime,
and then translated onto the CFT, using the AdS/CFT
correspondence. It is quasinormal frequencies that give us the
thermalization timescale which is very hard to compute directly.
Therefore, many authors have delved into studies of QNMs
associated with scalar, electromagnetic and gravitational
perturbations in various asymptotically AdS black holes
\cite{Chan,Birmingham,Cardoso, Konoplya,
Starinets,Kurita,Horowitz, Cardoso1, Cardoso2, Konoplya1}.

Although much attention has been paid to the study of the QNMs for
scalar, gravitational, electromagnetic field perturbations in the
SAdS and the Reissner-Norstr\"om AdS black-hole backgrounds,
however, to my best knowledge, at the moment the study of the QNMs
for Dirac field perturbation in asymptotically AdS spacetime is
still an open question. The main purpose of this paper is to study
the QNMs associated with the decay of the Dirac field perturbation
around a SAdS black hole.

The organization of this paper is as follows. In Sec. 2 the
decoupled Dirac equations and corresponding wave equations in the
SAdS spacetime are obtained by using Newman-Penrose formalism. In
Sec. 3 the numerical approach to computing the Dirac QNMs is
introduced. In Sec. 4 the numerical results for the Dirac QNMs in
the SAdS black hole are presented. Sec. 5 is devoted to a summary.
In appendix we present the relations between our wave-equation and
the standard one.

\section{Dirac equations in the Schwarzschild-anti-de Sitter spacetime}

We begin with the SAdS metric
\begin{eqnarray} \label{metric}
ds^2=f dt^2-\frac{1}{f}dr^2-r^2(d\theta^2+sin^2\theta d\varphi^2),
\end{eqnarray}
with
\begin{eqnarray}
f=1-\frac{2 M}{r}+\frac{r^2}{R^2},
\end{eqnarray}
where M represents the black hole mass, and R is the anti-de
Sitter radius. Hereafter we will take $R=1$ and measure everything
in terms of $R$. The black hole horizon is located at $r_+$, the
largest root of the function $f$, and its Hawking temperature is
given by
\begin{eqnarray}
T_H=\frac{\kappa}{2\pi}=\frac{1}{4\pi}\left.\frac{d  f}{d
r}\right|_{r_+}=\frac{3 r_+^2+1}{4\pi r_+ },
\end{eqnarray}
where $\kappa$ is a surface gravity. The mass parameter $M$ can be
viewed as a function of the horizon radius
\begin{eqnarray}
M=\frac{r_+^3}{2}\left(1+\frac{1}{r_+^2}\right).
\end{eqnarray}

The Dirac equations \cite{Page} are
\begin{eqnarray}
   &&\sqrt{2}\nabla_{BB'}P^B+i\mu \bar{Q}_{B'}=0, \nonumber \\
   &&\sqrt{2}\nabla_{BB'}Q^B+i\mu \bar{P}_{B'}=0,
\end{eqnarray}
where $\nabla_{BB'}$ is covariant differentiation, $P^B$ and $Q^B$
are the two-component spinors representing the wave function, $
\bar{P}_{B'}$ is the complex conjugate of $P_{B}$, and $\mu $ is
the particle mass. In the Newman-Penrose formalism \cite{Newman}
the equations become
\begin{eqnarray}\label{np}
   &&(D+\epsilon-\rho )P^0+
   (\bar{\delta}+\pi-\alpha )P^1=2^{-1/2}i\mu  \bar{Q}^{1'},\nonumber \\
   &&(\triangle+\mu -\gamma )P^1+
   (\delta+\beta -\tau )P^0=-2^{-1/2}i\mu
    \bar{Q}^{0'}, \nonumber\\
   &&(D+\bar{\epsilon}-\bar{\rho} )\bar{Q}^{0'}+
   (\delta+\bar{\pi}-\bar{\alpha} )\bar{Q}^{1'}=-2^{-1/2}i\mu  P^{1},\nonumber \\
  &&(\triangle+\bar{\mu} -\bar{\gamma} )\bar{Q}^{1'}+
   (\bar{\delta}+\bar{\beta} -\bar{\tau} )\bar{Q}^{0'}=2^{-1/2}i\mu
   P^{0}.
\end{eqnarray}
For the SAdS spacetime (\ref{metric}) the null tetrad can be taken
as
\begin{eqnarray}
  &&l^\mu=(\frac{r^2}{\Delta}, ~1, ~0, ~0 ), \nonumber \\
  &&n^\mu=\frac{1}{2}(1, ~-\frac{\Delta}{r^2}, ~0, ~0)\nonumber \\
  &&m^\mu=\frac{1}{\sqrt{2} r}\left(0, ~0, ~1, \frac{i}{sin\theta}\right),
\end{eqnarray}
with
\begin{eqnarray}
\Delta=r^2-2M r+r^4.
\end{eqnarray}
Then, if we set
\begin{eqnarray}
&&P^0=\frac{1}{r}f_1(r,\theta)e^{-i(\omega t-m\varphi)}, \nonumber \\
&&P^1=f_2(r,\theta)e^{-i(\omega t-m\varphi)}, \nonumber \\
&&\bar{Q}^{1'}=g_1(r,\theta)e^{-i(\omega t-m\varphi)}, \nonumber \\
&&\bar{Q}^{0'}=-\frac{1}{r}g_2(r,\theta)e^{-i(\omega t-m\varphi)},
\end{eqnarray}
where $\omega$ and $m$ are the energy and angular momentum of the
Dirac particles, Eq. (\ref{np}) can be simplified as
\begin{eqnarray} \label{DD}
&&{\mathcal{D}}_0 f_1+\frac{1}{\sqrt{2}}{\mathcal{L}}_{1/2}
f_2=\frac{1}{\sqrt{2}}i\mu r g_1,
\nonumber \\
&&{\Delta \mathcal{D}}_{1/2}^{\dag}
f_2-\sqrt{2}{\mathcal{L}}_{1/2}^{\dag} f_1=-\sqrt{2}i\mu r g_1,
\nonumber \\
&&{\mathcal{D}}_0 g_2-\frac{1}{\sqrt{2}}{\mathcal{L}}_{1/2}^{\dag}
g_1=\frac{1}{\sqrt{2}}i\mu r f_2,
\nonumber \\
&& {\Delta \mathcal{D}}_{1/2}^{\dag}
g_1+\sqrt{2}{\mathcal{L}}_{1/2} g_2=-\sqrt{2}i\mu r f_1,
\end{eqnarray}
with
 \begin{eqnarray}
 &&{\mathcal{D}}_n=\frac{\partial}{\partial r}-\frac{i K}
 {\Delta}+\frac{n}{\Delta}\frac{d \Delta}{d r},\nonumber \\
 &&{\mathcal{D}}^{\dag}_n=\frac{\partial}{\partial r}+\frac{i K}
 {\Delta}+\frac{n}{\Delta}\frac{d \Delta}{d r},\nonumber \\
 &&{\mathcal{L}}_n=\frac{\partial}{\partial \theta}+\frac{m}{\sin \theta }
 +n\cot \theta,\nonumber \\
 &&{\mathcal{L}}^{\dag}_n=\frac{\partial}{\partial \theta}-\frac{m}{\sin \theta }
 +n\cot \theta, \nonumber \\
 &&K=r^2\omega.\label{ld}
 \end{eqnarray}
It is now apparent that the variables can be separated by the
substitutions
 \begin{eqnarray}
f_1={\mathbb{R}}_{-1/2}(r)S_{-1/2}(\theta),\nonumber \\
f_2={\mathbb{R}}_{+1/2}(r)S_{+1/2}(\theta),\nonumber \\
g_1={\mathbb{R}}_{+1/2}(r)S_{-1/2}(\theta),\nonumber \\
g_2={\mathbb{R}}_{-1/2}(r)S_{+1/2}(\theta).
 \end{eqnarray}
Then, Eq. (\ref{DD}) reduces to the following radial and angular
parts
\begin{eqnarray}\label{dd2}
&&\sqrt{\Delta}{\mathcal{D}}_0 {\mathbb{R}}_{-1/2}=(\lambda+i\mu
r)\sqrt{\Delta} {\mathbb{R}}_{+1/2}, \\
\label{dd3}&&\sqrt{\Delta}{\mathcal{D}}_0^{\dag}
(\sqrt{\Delta}{\mathbb{R}}_{+1/2})=(\lambda-i\mu r) {\mathbb{R}}_{-1/2},\\
&&{\mathcal{L}}_{1/2} S_{+1/2}=-\lambda S_{-1/2}, \label{aa1}\\
&&{\mathcal{L}}_{1/2}^{\dag} S_{-1/2}=\lambda S_{+1/2}.\label{aa2}
\end{eqnarray}
We can eliminate $S_{+1/2}$ (or $S_{-1/2}$) from Eqs. (\ref{aa1})
and (\ref{aa2}) and obtain
\begin{eqnarray}
&&{\mathcal{L}}_{1/2}^{\dag}{\mathcal{L}}_{1/2}
S_{+1/2}=-\lambda^2 S_{+1/2}, \\
&&{\mathcal{L}}_{1/2}{\mathcal{L}}_{1/2}^{\dag}
S_{-1/2}=-\lambda^2 S_{-1/2}.
\end{eqnarray}
Both them can be expressed as
\begin{eqnarray}\label{ang}
\left[\frac{1}{sin\theta}\frac{d}{d
\theta}\left(sin\theta\frac{d}{d\theta}\right)-\frac{m^2+2mscos\theta+s^2cos^2\theta}
{sin^2\theta}+s+A_s\right]S_s=0,
\end{eqnarray}
here and hereafter we take $s=+1/2$ for the case $S_{+1/2}$
(${\mathbb{R}}_{+1/2}$) and $s=-1/2$ for $S_{-1/2}$ (
${\mathbb{R}}_{- 1/2}$), and $A_{+1/2}=\lambda^2-2s$ and
$A_{-1/2}=\lambda^2$.  The angular equation (\ref{ang}) can be
solved exactly and $A_s=(l-s)(l+s+1)$, where $l$ is the quantum
number characterizing the angular distribution. So, we have
$\lambda^2=\left(l+\frac{1}{2}\right)^2$ for both cases $s=+1/2$
and $s=-1/2$.

In this paper we will focus our attention on the massless case.
Then, we can eliminate ${\mathbb{R}}_{-1/2}$ (or
$\sqrt{\Delta}{\mathbb{R}}_{+1/2}$) from Eqs. (\ref{dd2}) and
(\ref{dd3}) to obtain a radial decoupled Dirac equation for
$\sqrt{\Delta} {\mathbb{R}}_{+1/2}$ (or ${\mathbb{R}}_{-1/2}$),
and we find that both them can be expressed as
\begin{eqnarray}  \label{T1}
&& \Delta^{-s} \frac{d }{d r}\left(\Delta^{1+s}
\frac{d{\mathbb{R}}_{s}}{d r}\right)+P{\mathbb{R}}_{s}
 =0,
 \end{eqnarray}
with
 \begin{eqnarray}
&&P=\frac{K^2-is K\frac{d \Delta}{dr}}{ \Delta} +4s i  \omega
r+\frac{1}{2}\left(s+\frac{1}{2}\right)\frac{d^2 \Delta}{d r^2}
-\lambda^2,
 \end{eqnarray}
We now try to express Eq. (\ref{T1}) as a wave-equation.
Introducing an usual tortoise coordinate
 $
dr_*=(r^2/\Delta) dr $
 and resolving the equation in the form
 \begin{eqnarray} \label{Rphi}
&&{\mathbb{R}}_{s}=\frac{\Delta^{-s/2}}{r} \Psi_s,
 \end{eqnarray}
then we obtain
\begin{eqnarray}
&&  \frac{d^2 \Psi_s}{d r_*^2}+\left\{\frac{d H}{d
r_*}-H^2+\frac{\Delta}{r^4}P \right\}\Psi_s
 =0,\label{LV1}
 \end{eqnarray}
where
 \begin{eqnarray}
H=-\left[\frac{s}{2r^2}\frac{d\Delta}{dr}+\frac{\Delta}{r^3}
\right].
 \end{eqnarray}
We find that Eq. (\ref{LV1}) can be expressed as
\begin{eqnarray}\label{wave}
\frac{d \Psi_s }{d r_*^2}+(\omega ^2-V )\Psi_s =0,
\end{eqnarray}
where
\begin{eqnarray}\label{Poten}
V=-\frac{\Delta}{4 r^2}\frac{d}{d
r}\left[r^2\frac{d}{dr}\left(\frac{\Delta}{r^4}\right)\right]+\frac{s^2
r^4}{4}\left[\frac{d}{d
r}\left(\frac{\Delta}{r^4}\right)\right]^2+is \omega r^2\frac{d}{d
r}\left(\frac{\Delta}{r^4}\right)+\frac{\lambda^2 \Delta}{r^4}.
\end{eqnarray}
The potential $V$ is complex and the complex frequency $\omega$ is
not separated from $V$. We note that potentials which are
functions of complex frequency have always been used to study QNMs
in Refs. \cite{Leaver} \cite{Kokkotas}\cite{Cho}\cite{Simone}
because the key point to investigate QNMs is that we should get
solutions of the original equations with appropriate boundary
conditions presented in the introduction.

We should point out here that, for the Dirac fields, we can also
obtain another two standard wave-equations $\left(\frac{d^2}{d
r_*^2}+\omega^2\right)Z_{\pm}=V_{\pm}Z_{\pm}$ with the potentials
$V_{\pm}=\lambda^2\frac{ \Delta }{r^4}\pm \lambda \frac{d}{d
r_*}\frac{\sqrt{\Delta}}{r^2}$ by using similar approach in Refs.
\cite{Cho}\cite{Jing1}\cite{Jing2}. At first sight the potentials
are simpler than Eq. (\ref{Poten}). However, if the Dirac
quasinormal frequencies of the SAdS black hole are studied by
using the potentials $V_{\pm}$ and Horowitz-Hubeny
approach\cite{Horowitz}, we find that we can not expand the
potentials about the event horizon $r_+$ due to the potentials
include a non-polynomial factor $\sqrt{\Delta}$. Although we can
expand them about $y=0$ with $y^2=r-r_+$, the expansion function
will be an infinite series. Therefore, the numerical calculation
will take very long computer time to get Dirac fundamental
quasinormal frequencies, and it is very difficult to find higher
overtone modes. On the other hand, in appendix we present the
relations between the two kinds of the wave functions in our
wave-equation and the standard one which imply that the
quasinormal frequencies are the same for both cases. These are the
reasons why we use the potential (\ref{Poten}) to study Dirac QNMs
of the SAdS black hole here.

\section{Numerical Approach to Computing Dirac Quasinormal Modes}

For the SAdS black hole, the potential (\ref{Poten}) diverges at
infinity, so we must require that $\Psi_s$ vanish there. QNMs are
defined to be modes with only ingoing waves near the event
horizon. The boundary conditions on wave function $\Psi_s$ (or
${\mathbb{R}}_{s}$) at the horizon $(r=r_+)$ and infinity
$(r\rightarrow +\infty)$ can be expressed mathematically as
\begin{eqnarray} \label{Bon}
\Psi_s \sim \Delta^{s/2} {\mathbb{R}}_{s} \sim \left\{
\begin{array}{ll} \Delta^{-s/2}e^{-i\omega r_*} &
r\rightarrow r_+, \\
     0 &      r\rightarrow +\infty.
\end{array} \right.
\end{eqnarray}
Equations (\ref{wave}), (\ref{Poten}) and (\ref{Bon}) determine an
eigenvalue problem for the quasinormal frequency $\omega$ of the
Dirac field perturbation.

In what follows, we will calculate the quasinormal frequencies for
outgoing Dirac field (i.e., for case $s=-1/2$ \cite{Leaver}) by
using the Horowitz-Hubeny approach \cite{Horowitz}. Writing
$\Phi_s$ for a generic wave function as
\begin{eqnarray}
\Phi_s=\Psi_s e^{i\omega r_*},
\end{eqnarray}
we find that Eq. (\ref{wave}) can be rewritten as
\begin{eqnarray}\label{wave1}
f^2 \frac{d^2 \Phi_s}{d r^2}+\left(f\frac{d f}{d r}-2i\omega
f\right)\frac{d \Phi_s}{d r}-V\Phi_s=0.
\end{eqnarray}
In order to map the entire region of interest, $r_+<r<+\infty$,
into a finite parameter range, we change variable to $x=1/r$.
Define a new function $B(x)$ as
\begin{eqnarray}
B(x)=x^2-\frac{1+x_+^2}{x_+^3}x^3+1,
\end{eqnarray}
then $f$ can be rescaled as
\begin{eqnarray}
f=x^2\Delta=\frac{B(x)}{x^2}.
\end{eqnarray}
In terms of the new variable $x$, Eq. (\ref{wave1}) can be
expressed as
\begin{eqnarray} \label{wave2}
S(x)\frac{d^2 \Phi_s}{d x^2}+\frac{T(x)}{x-x_+}\frac{d \Phi_s}{d
x}+\frac{U(x)}{(x-x_+)^2}\Phi_s=0,
\end{eqnarray}
where the coefficient functions are described as
\begin{eqnarray}
S(x)&=&\frac{B(x)^2}{(x-x_+)^2}=\left(\frac{1+x_+^2}
{x_+^3}x^2+\frac{x}{x_+^2}+\frac{1}{x_+}\right)^2, \nonumber \\
T(x)&=&\frac{B(x)}{x-x_+}\left(\frac{d B(x)}{d
x}+2I\omega\right)=\left(\frac{1+x_+^2}
{x_+^3}x^2+\frac{x}{x_+^2}+\frac{1}{x_+}\right)
\left(\frac{3(1+x_+^2)}{x_+^3}x^2-2x-2I\omega\right), \nonumber \\
U(x)&=&-V=\frac{B(x)}{4}\frac{d^2 B(x)}{d
x^2}-\frac{s^2}{4}\left(\frac{d B(x)}{d x}\right)^2+is\omega
\frac{d B(x)}{d x}-\lambda^2B(x).
\end{eqnarray}
Since $S(x)$, $T(x)$ and $U(x)$ are all polynomial of degree 4, we
can expand them about the horizon $x=x_+$ as
$S(x)=\sum_{n=0}^{4}S_n(x-x_+)^n$, and similarly for $T(x)$ and
$U(x)$.

To evaluate quasinormal frequencies by using Horowitz-Hubeny
method, we need to expand the solution to the wave function
$\Phi_s$ around $x_+$,
\begin{eqnarray}\label{exp}
\Phi_s=(x-x_+)^\alpha \sum_{k=0}^{\infty}a_k (x-x_+)^k,
\end{eqnarray}
and to find the roots of the equation $\Phi_s|_{x=0}=0$. First, we
should determine the behavior of the solutions near the black hole
horizon. Then, we should substitute Eq. (\ref{exp}) into the
differential equation (\ref{wave2}) in order to get a recursion
relation for $a_k$. At last, we have to look for some true root
which is the sought quasinormal frequency.

To find index $\alpha$ we set $\Phi_s=(x-x_+)^\alpha$ near the
event horizon and substitute it into Eq. (\ref{wave2}). Then, we
have
\begin{eqnarray}\label{ind}
\alpha (\alpha -1)S_0+\alpha T_0+U_0=4\alpha \kappa (\alpha
\kappa-i\omega)-2 s \kappa \left(\frac{s}{2}\kappa+i\omega
\right)=0,
\end{eqnarray}
which has two solutions $\alpha=-\frac{s}{2}$ and
$\alpha=\frac{s}{2}+\frac{i\omega}{\kappa}$. Since we need only
ingoing modes near the event horizon, that is to say, $\Phi_s$
must satisfy the boundary condition (\ref{Bon}),  we take
$\alpha=-\frac{s}{2}$. Then $\Phi_s$ is described by
\begin{eqnarray}\label{exp1}
\Phi_s=(x-x_+)^{-s/2} \sum_{k=0}^{\infty}a_k (x-x_+)^k.
\end{eqnarray}
Substituting Eq. (\ref{exp1}) into Eq. (\ref{wave2}) we find the
following recursion relation for  $a_n$
\begin{eqnarray}\label{an}
a_n=-\frac{1}{Z_n}\sum_{k=0}^{n-1}\left[\left(k-\frac{s}{2}\right)
\left(k-\frac{s}{2}-1\right)S_{n-k}+\left(k-\frac{s}{2}\right)
T_{n-k}+U_{n-k}\right]a_k,
\end{eqnarray}
where
\begin{eqnarray}
Z_n=4n\kappa [(n-s)\kappa -i \omega].
\end{eqnarray}
The boundary condition (\ref{Bon}) at $r\rightarrow \infty$ is now
becomes
\begin{eqnarray}\label{Poo}
\Phi_s=(x-x_+)^{-s/2} \sum_{k=0}^{\infty}a_k (x-x_+)^k\rightarrow
0,
\end{eqnarray}
as $x\rightarrow 0$. Taking the limit of the equation we get
\begin{eqnarray}\label{po}
\sum_{k=0}^{\infty}a_k (-x_+)^{k}=0.
\end{eqnarray}
The computation of quasinormal frequencies is now reduced to that
of finding a numerical solution of the Eq. (\ref{po}).  We first
truncate the sum (\ref{po}) at some large $k=N$ and then check
that for greater $k$ (say $40$ larger than $N$) the roots converge
to some true roots, i.e., quasinormal frequencies. In the series
(\ref{an}) and (\ref{po}) each next term depends on all the
preceding terms through the recursion relations, and the roots of
(\ref{po}) suffer a sharp change for a small change on any of the
input parameters, especially for finding higher overtone modes. In
order to avoid the ``noisy" we increase the precision of all the
input data and we retain 50-digital precision in all the
intermediate process. We find that the 50-digital precision is
necessary because we have checked that, for $r_+=100$ and $N=400$,
20-digital precision gives 4 convergent roots, 40-digital
precision shows 9 convergent roots and 50-digital precision
presents 11 convergent roots.

\vspace*{1.0cm}

\section{Numerical results for Dirac Quasinormal Frequencies}

In this section, we will represent the numerical results obtained
by using the numerical procedure just outlined in the previous
section. The results will be organized into four subsections: the
fundamental QNMs (the overtone number $n=0$ and the size of black
hole run from $r_+=0.4$ to $r_+=100$), large black holes
($r_+=1000$ and $r_+=100$),  an intermediate black hole ($r_+=1$),
and  a small black hole ($r_+=0.4$) QNMs (the overtone number runs
from $n=0$ to $n=10$).

\subsection{Fundamental Quasinormal Modes}

\begin{figure}
\includegraphics[scale=0.55]{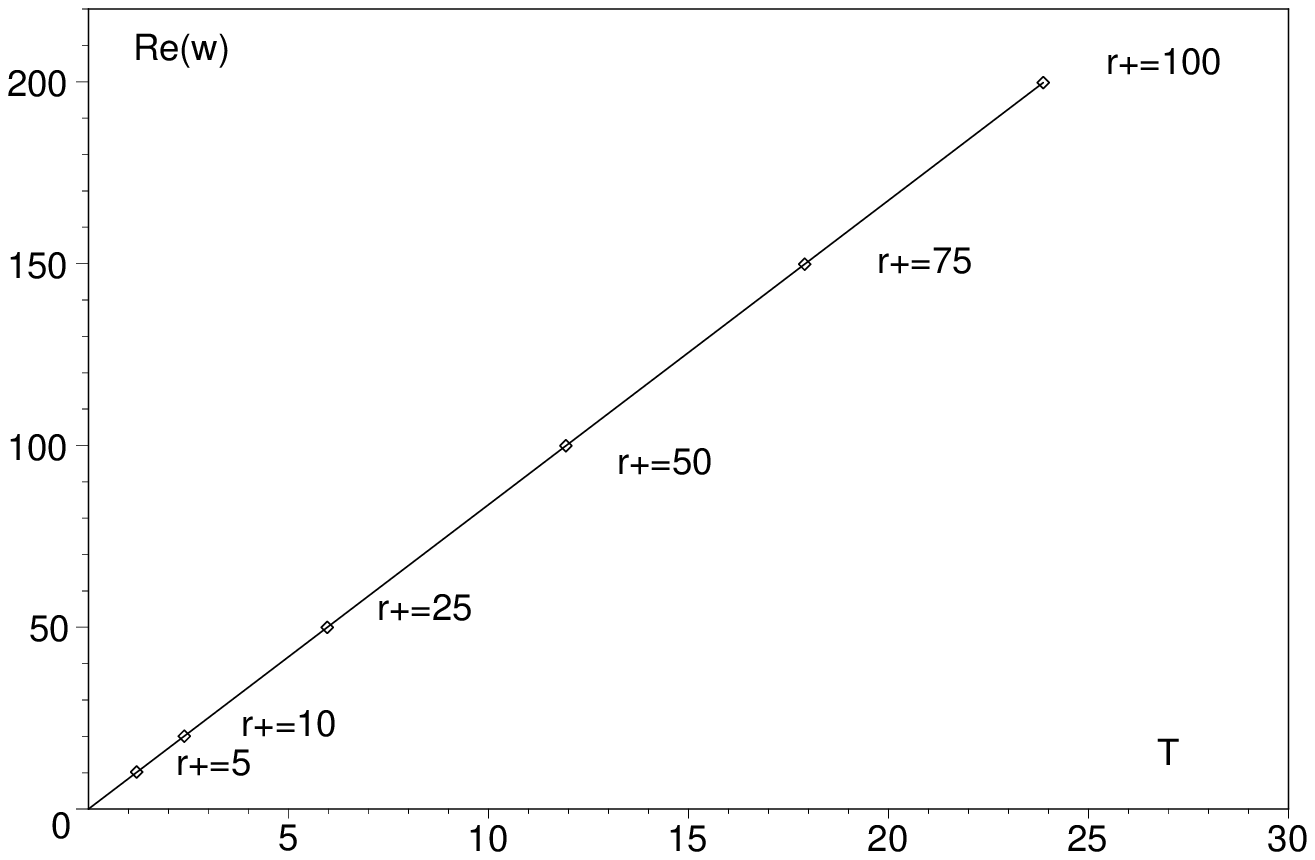}\hspace{0.5cm}%
\includegraphics[scale=0.55]{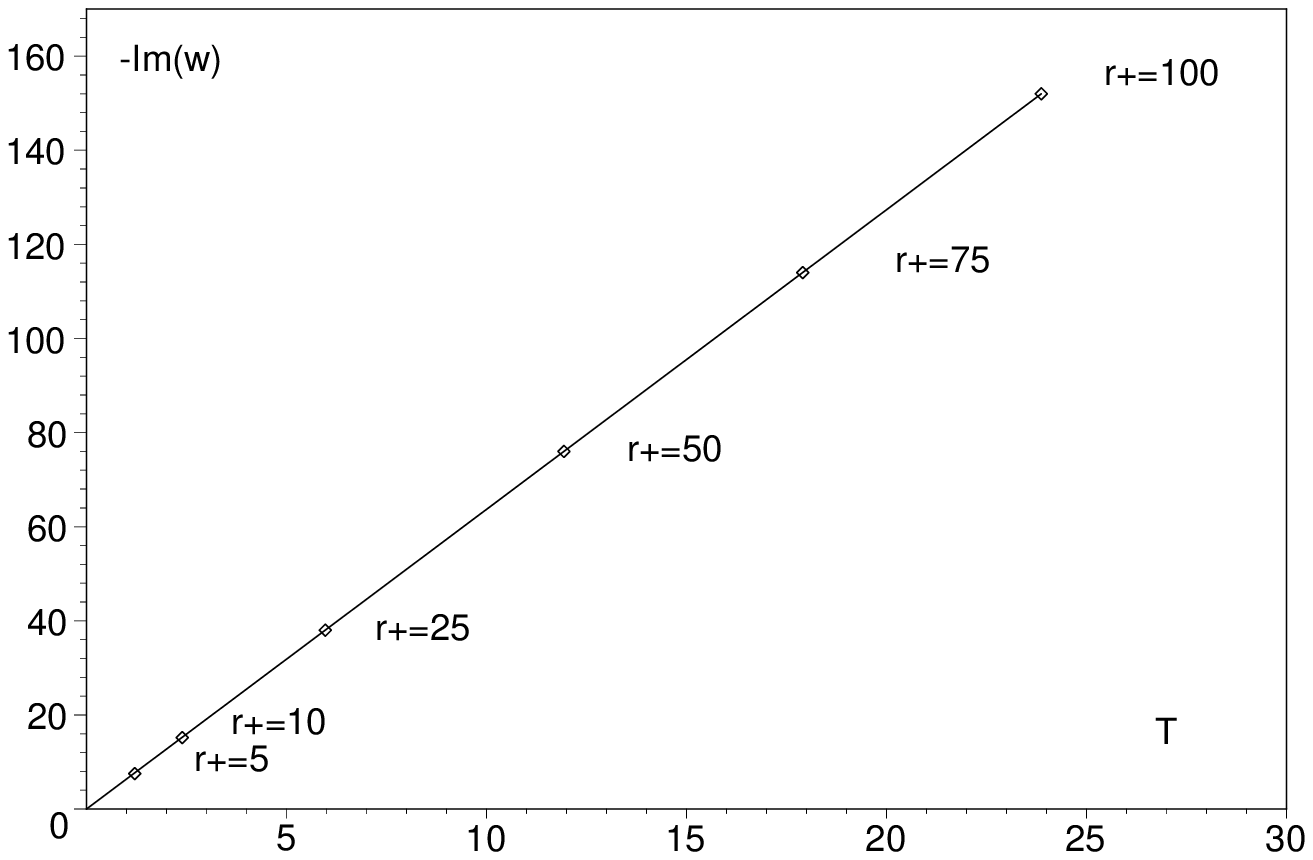}
\caption{\label{fig1}Graphs of fundamental quasinormal frequencies
$\omega$ versus Hawking temperature $T$  for large black holes.
The left figure is drawn for the cases $Re(\omega)$ and right one
for $Im(\omega)$. The figures show that both the real and the
imaginary parts of the frequency are linear function of $T$, and
the lines are described by $Re(\omega)=8.367 T $ and
$-Im(\omega)=6.371 T $. }
\end{figure}

\begin{figure}
\includegraphics[scale=0.55]{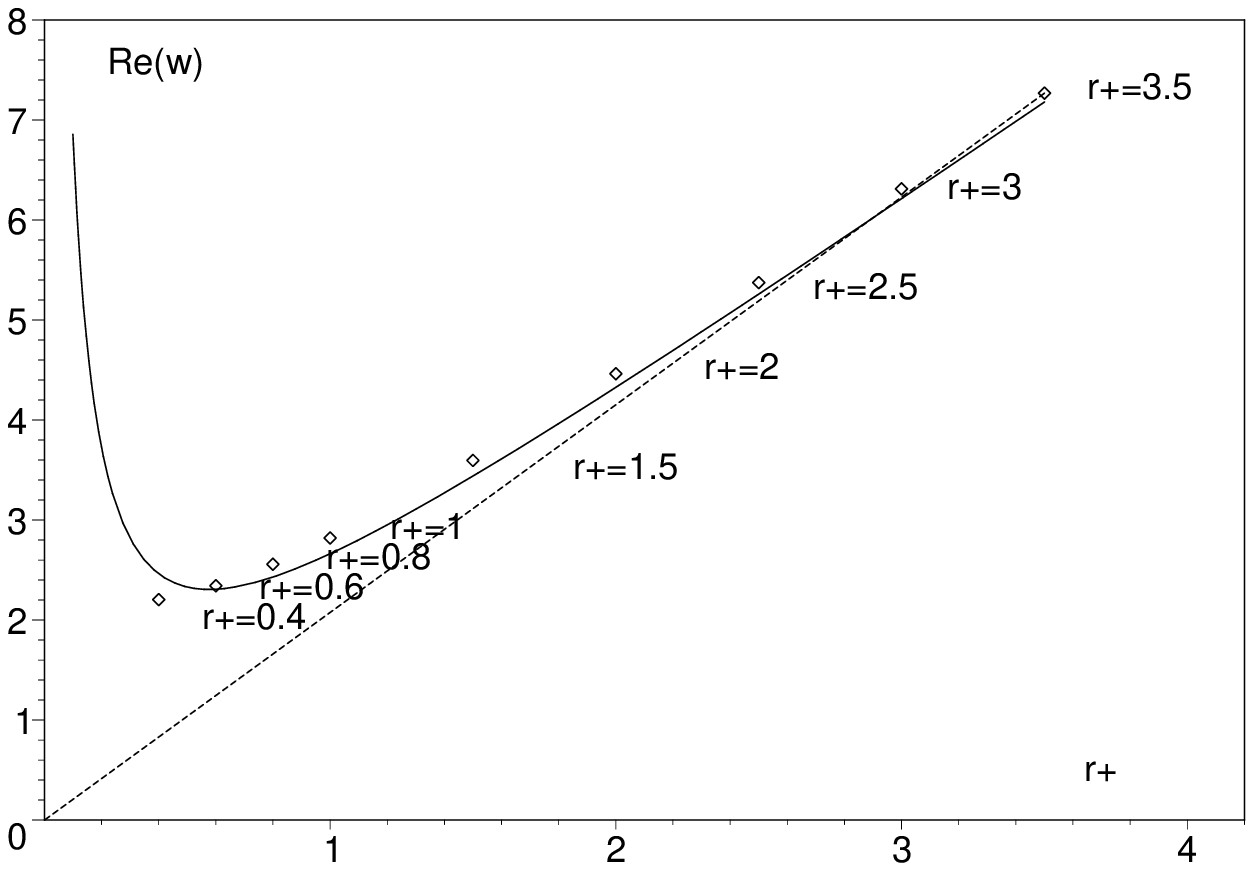}\hspace{0.5cm}%
\includegraphics[scale=0.55]{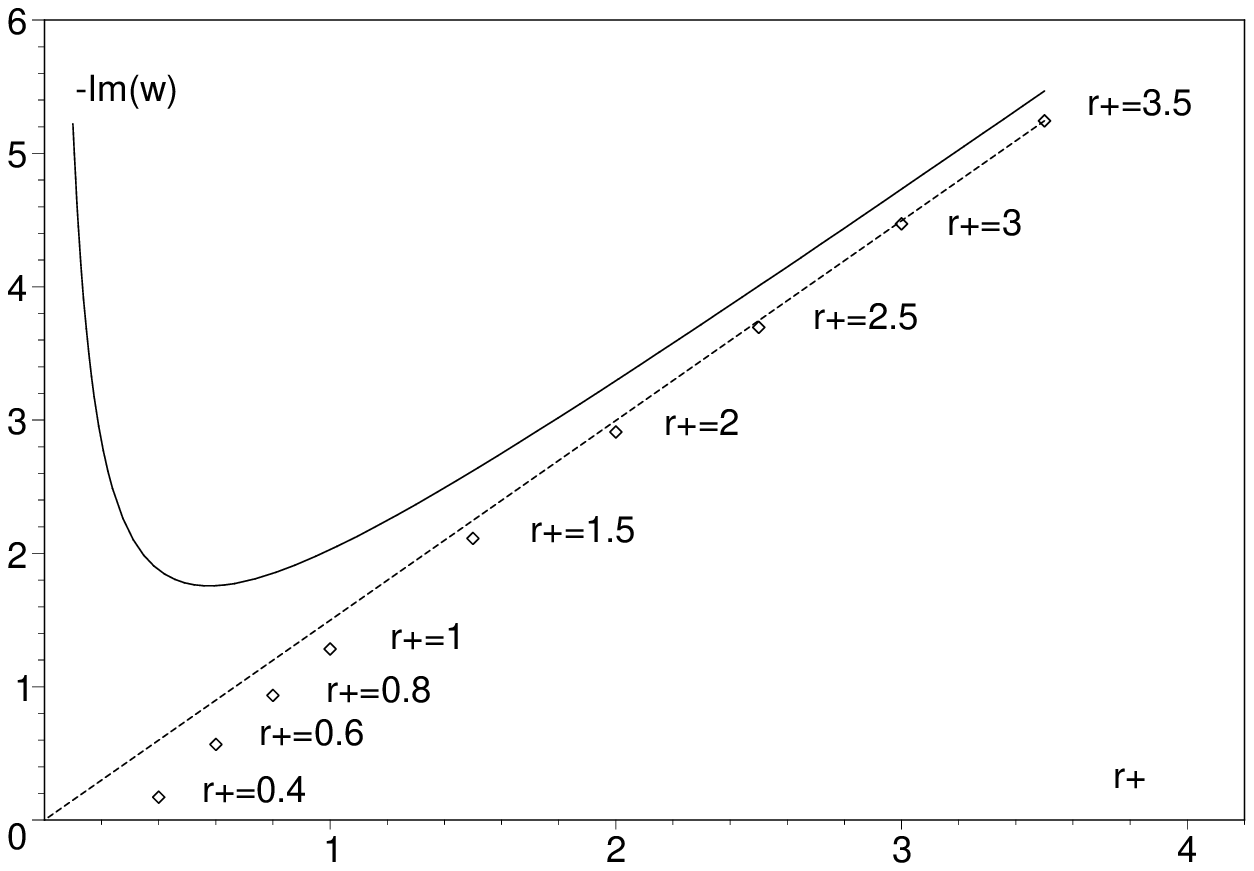}
\caption{\label{fig2}$Re(\omega)$ and  $Im(\omega)$ of the
fundamental quasinormal frequencies versus radius $r_+$ for
intermediate and small black holes.  The dashed lines are
$Re(\omega)=2.706 r_+ $ in the left figure and $-Im(\omega)=1.499
r_+ $ in the right one, and the solid lines are $Re(\omega)=8.367
T $ in the left figure and $-Im(\omega)=6.371 T $ in the right
one.}
\end{figure}

The fundamental quasinormal frequencies ($n=0$) corresponding to
$\lambda=1$ Dirac perturbation of the SAdS black hole are given by
table (\ref{table1}) and corresponding results are drawn in Fig.
(\ref{fig1}) for large black holes and in Fig. (\ref{fig2}) for
intermediate and small black holes. From the table and figures we
find that, for large black holes, both the real and the imaginary
parts of the quasinormal frequency are linear functions of the
Hawking temperature, and the lines are described by
\begin{eqnarray}\label{large}
&& Re(\omega)=8.367 T, \\
&-&Im(\omega)=6.371 T.
\end{eqnarray}
The Fig. (\ref{fig2}) shows that, for intermediate and small size
black holes, the quasinormal frequencies do not scale with the
Hawking temperature. Here $Re(\omega)$ approximates the
temperature $T$ more closely than the black hole size $r_+$, but
it is not diverging for small black holes. However, for
$Im(\omega)$, the points continue to lie along (the points for
black holes with radius $r_+\leq 1.5$ lie slightly off) a straight
line
\begin{eqnarray}
-Im(\omega)=1.499 r_+ . \end{eqnarray}

\begin{table}
\caption{\label{table1} The fundamental quasinormal frequencies
($n=0$) corresponding to $\lambda=1$ Dirac perturbation of the
SAdS black hole.}
\begin{tabular}{c|c|c||c|c|c}
\hline \hline $~~~~~~ r_+ ~~~~~~$ & ~~~~~~Re($\omega$)~~~~~~ &
~~~~~~ -Im($\omega$) ~~~~~~& $~~~~~~ r_+ ~~~~~~$ &
~~~~~~Re($\omega$)~~~~~~ & ~~~~~~-Im($\omega$) ~~~~~~ \\
\hline
100 & 199.769  & 152.016     &  2.5 &  5.3747  &  3.6959   \\
75  & 149.834  & 114.012    &  2   &  4.4620  &  2.9113   \\
50  & 99.8993  & 76.0043      &  1.5 &  3.5961  &  2.1126   \\
25  & 49.9795  & 37.9952      &  1   &  2.8219  &  1.2836   \\
10  & 20.0744  & 15.1751      &  0.8 &  2.5596  &  0.9348   \\
5   &  0.1834  &  7.5476      &  0.6 &  2.3439  &  0.5677   \\
3.5 &  7.2683  &  5.2449      &  0.4 &  2.2045  &  0.1714    \\
3   &  6.3135  &  4.4727   & & &  \\
\hline \hline
\end{tabular}
\end{table}
\begin{figure}
\includegraphics[scale=0.55]{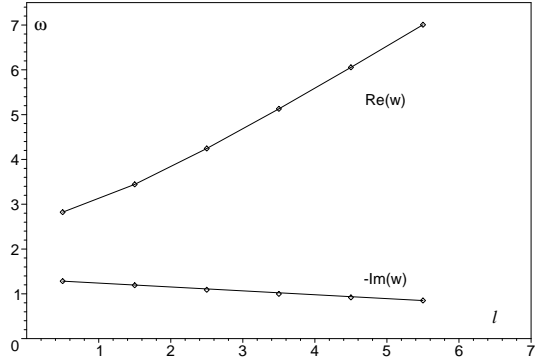}
\caption{\label{fig3}Dependence of the fundamental quasinormal
frequencies $\omega $ on the angular quantum number $l $ for an
intermediate SAdS black hole ($r_+=1$), which shows that  $l$ has
the surprising effect of decreasing $|Im(\omega)|$ and increasing
$Re(\omega)$. }
\end{figure}

We also study the relation between the quasinormal frequencies and
the angular quantum number and find that, for large black hole
(say $r_+=100$), the quasinormal frequencies depend very weakly on
the low angular quantum number (say $l=3/2$ and $l=5/2$). However,
for intermediate black hole ($r_+=1$), Fig (\ref{fig3}) shows that
the angular quantum number $l$ mode has the surprising effect of
decreasing $|Im(\omega)|$ (increasing the damping time scale) and
increasing $Re(\omega)$ (decreasing the oscillation time scale).
We have checked that the result for the intermediate black hole is
also valid for the small black holes.


\subsection{ Quasinormal Modes for Large black holes}

\begin{figure}
\includegraphics[scale=0.55]{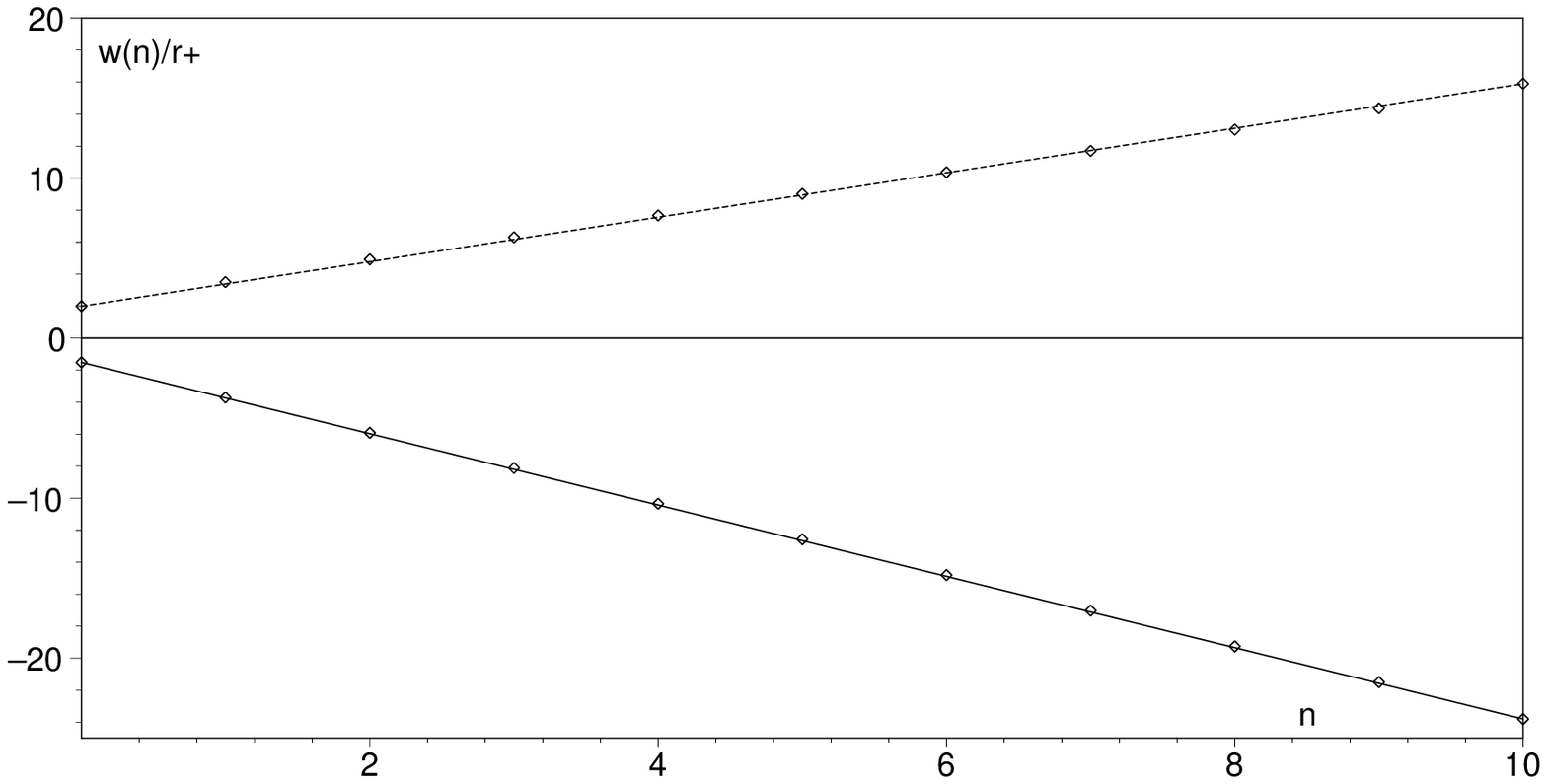}
\caption{\label{fig4}$\frac{\omega(n)}{r_+}$ versus overtone
number $n$ for large SAdS black holes.  The dashed line describes
$\frac{Re(\omega)(n)}{r_+} $ and the solid line is
$\frac{Im(\omega)(n)}{r_+}$. The behave of the QNMs can be
expressed as $\frac{\omega(n)}{r_+}\sim (1.385-2.220 i)n+2.04-1.50
i$.}
\end{figure}

We show in Table (\ref{table2}) and Fig (\ref{fig4}) the
quasinormal frequencies corresponding to $\lambda=1$ Dirac
perturbation of large SAdS black holes ($r_+=1000$ and $r_+=100$).
Perhaps the most interesting result for these large black holes is
that the points for each overtone modes in Fig. (\ref{fig4})
continue to lie along the straight lines
$\frac{Re(\omega(n))}{r_+}=1.35 n +2.136$ and
$\frac{Im(\omega(n))}{r_+}=-(2.25  n +1.50)  $. We have checked
that the behave of the QNMs can be described as
\begin{eqnarray}
\frac{\omega(n)}{r_+}\sim (1.35-2.25 i) n +(2.136-1.500 i)
\end{eqnarray}
with an error of about $2\% $. This leads to the spacing
\begin{eqnarray}
\frac{\omega_{(n+1)}-\omega_{(n)}}{r_+}=(1.35-2.25 i).
\end{eqnarray}

\begin{table}
\caption{\label{table2} Quasinormal frequencies corresponding to
$\lambda=1$ Dirac perturbation of large SAdS black holes
($r_+=1000$, $r_+=100$).}
\begin{tabular}{c|c|c||c|c}
 \hline \hline
\multicolumn{3}{c||} {$r_+=1000$  } & \multicolumn{2}{c} {$r_+=100$  }\\
 \hline  $~~~~~~ n ~~~~~~$ & ~~~~~~Re($\omega$)~~~~~~ & ~~~~~~
-Im($\omega$) ~~~~~~ &
~~~~~~Re($\omega$)~~~~~~ & ~~~~~~-Im($\omega$) ~~~~~~ \\
\hline
0  & 1997.59  & 1520.19         & 199.769   &  152.016    \\
1  & 3505.31  & 3705.82         & 350.548   &  370.580   \\
2  & 4923.53  & 5909.28         & 492.377   &  590.927   \\
3  & 6305.63  & 8123.53         & 630.595   &  812.353   \\
4  & 7668.35  & 10344.6         & 766.875   &  1034.46   \\
5  & 9019.10  & 12570.3         & 901.947   &  1257.03        \\
6   &  10361.4  &  14799.3      &  1036.20  &  1479.94   \\
7   &  11697.8  &  17030.9      &  1169.85  &  1703.10  \\
8   &  13029.8  &  19264.5      &  1303.05  &  1926.45    \\
9   &  14358.2  &  21499.7      &  1435.90  &  2149.98   \\
10  &  15896.7  &  23802.2      &  1601.11  &  2371.15  \\
 \hline \hline
\end{tabular}
\end{table}


\subsection{ Quasinormal Modes for intermediate size black hole}

\begin{figure}
\includegraphics[scale=0.55]{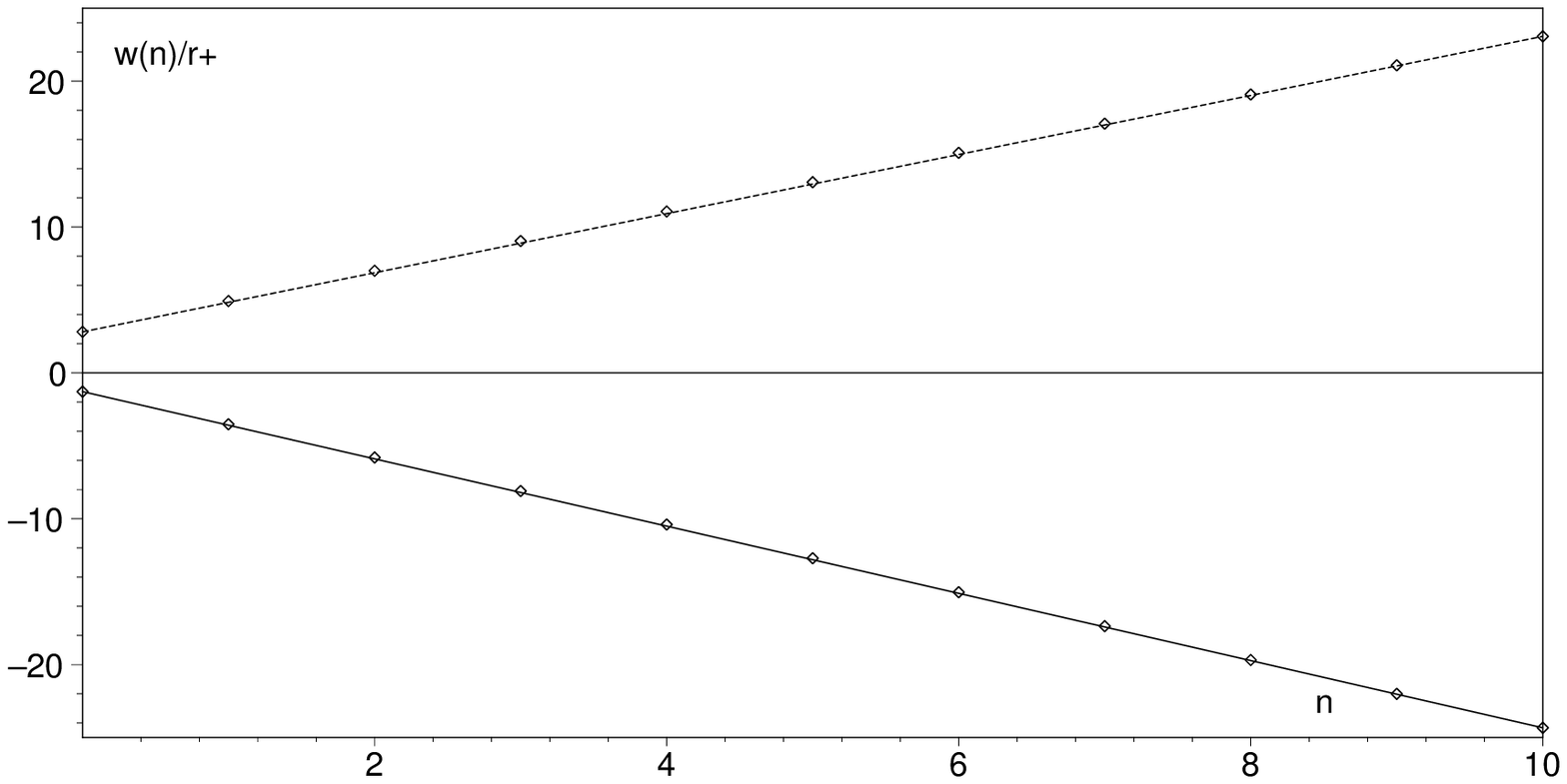}
\caption{\label{fig5}$\frac{\omega(n)}{r_+}$ versus overtone
number $n$ for an intermediate SAdS black hole ($r_+=1$). The
dashed line describes $\frac{Re(\omega)(n)}{r_+} $ and the solid
line is $\frac{Im(\omega)(n)}{r_+}$. The behave of the QNMs can be
expressed as $\frac{\omega(n)}{r_+}\sim (2.026-2.298 i)n+2.86-1.26
i$.}
\end{figure}

In Table (\ref{table3}) and Fig. (\ref{fig5}) we show some of the
quasinormal frequencies corresponding to $\lambda=1$ Dirac
perturbation of an intermediate black hole  ($r_+=1$). We find
again that the points for each overtone modes continue to lie
along the straight lines $\frac{Re(\omega(n))}{r_+}=2.02 n +2.86$
and $\frac{Im(\omega(n))}{r_+}=-(2.32  n +1.26)  $, i. e., the
behave of the QNMs can be expressed as
\begin{eqnarray}
\frac{\omega(n)}{r_+}\sim (2.02-2.32 i) n +(2.86-1.26 i)
\end{eqnarray} with an error of about $2\%$, and a spacing is given
by
\begin{eqnarray}
\frac{\omega_{(n+1)}-\omega_{(n)}}{r_+}=(2.02-2.32 i).
\end{eqnarray}

\begin{table}
\caption{\label{table3} Quasinormal frequencies corresponding to
$\lambda=1$ Dirac perturbation of an intermediate SAdS black hole
( $r_+=1$).}
\begin{tabular}{c|c|c||c|c|c}
\hline \hline $~~~~~~ n ~~~~~~$ & ~~~~~~Re($\omega$)~~~~~~ &
~~~~~~ -Im($\omega$) ~~~~~~& $~~~~~~ n ~~~~~~$ &
~~~~~~Re($\omega$)~~~~~~ & ~~~~~~-Im($\omega$) ~~~~~~ \\
\hline

0  & 2.82189  & 1.28361      &  6   &  15.0891  &  15.0280   \\
1  & 4.93748  & 3.51822      &  7   &  17.0922  &  17.3505   \\
2  & 7.00231  & 5.79451      &  8   &  19.0915  &  19.6761   \\
3  & 9.04161  & 8.08963      &  9   &  21.0877  &  22.0041   \\
4  & 11.0661  & 10.3958      &  10  &  23.0813  &  24.3341   \\
5  & 13.0810  & 12.7093      &    &    &     \\
 \hline \hline
  \end{tabular}
\end{table}

\subsection{ Quasinormal Modes for small black hole}

\begin{figure}
\includegraphics[scale=0.55]{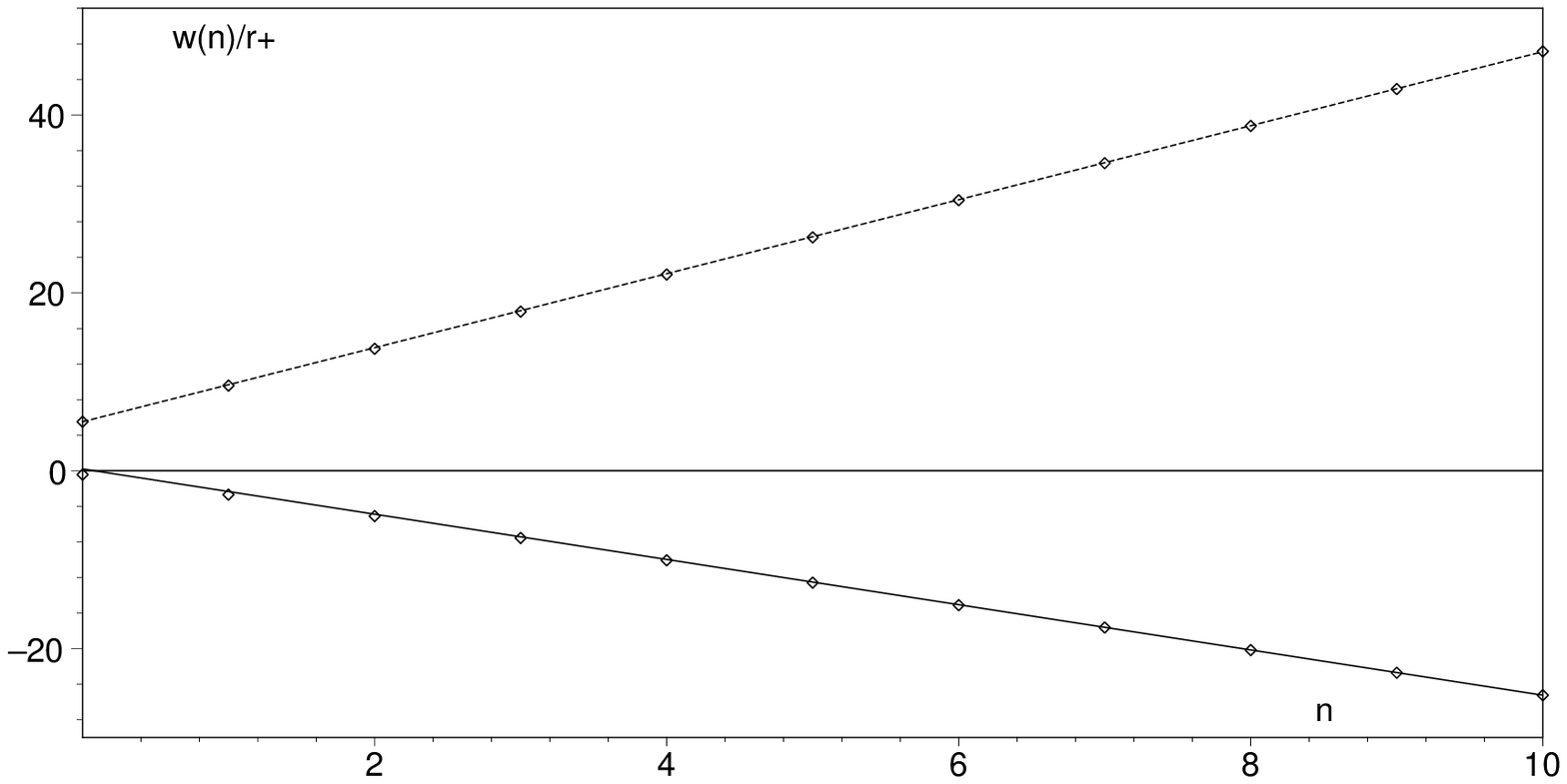}
\caption{\label{fig6}$\frac{\omega(n)}{r_+}$ versus overtone
number $n$ for a small SAdS black hole ($r_+=0.4$). The dashed
line describes $\frac{Re(\omega)(n)}{r_+} $ and the solid line is
$\frac{Im(\omega)(n)}{r_+}$. The behave of the QNMs can be
expressed as $\frac{\omega(n)}{r_+}\sim (4.18-2.48 i)n+5.45-0.20
i$.}
\end{figure}

We show in Table (\ref{table4}) and Fig. (\ref{fig6})  some of the
quasinormal frequencies corresponding to $\lambda=1$ Dirac
perturbation of a small black hole ($r_+=0.4$). It is shown that
the points for each overtone modes continue to lie along the
straight lines $\frac{Re(\omega(n))}{r_+}=4.18 n +5.45$ and
$\frac{Im(\omega(n))}{r_+}=-(2.48  n +0.20)  $. That is to say,
the behave of the QNMs can be written as
\begin{eqnarray}
\frac{\omega(n)}{r_+}\sim (4.18-2.48 i) n +(5.45-0.20 i)
\end{eqnarray}
with an error of about $1\%$ except $\omega(0)/r_+$, and a spacing
is shown by
\begin{eqnarray}
\frac{\omega_{(n+1)}-\omega_{(n)}}{r_+}=(4.18-2.48 i).
\end{eqnarray}

\begin{table}
\caption{\label{table4} Quasinormal frequencies corresponding to
$\lambda=1$ Dirac perturbation of a small SAdS black hole (
$r_+=0.4$).}
\begin{tabular}{c|c|c||c|c|c}
\hline \hline $~~~~~~ n ~~~~~~$ & ~~~~~~Re($\omega$)~~~~~~ &
~~~~~~ -Im($\omega$) ~~~~~~& $~~~~~~ n ~~~~~~$ &
~~~~~~Re($\omega$)~~~~~~ & ~~~~~~-Im($\omega$) ~~~~~~ \\
\hline

0  & 2.20450  &  0.17410      &  6   &  12.1661  &  6.03911   \\
1  & 3.82877  &  1.07155      &  7   &  13.8347  &  7.05100   \\
2  & 5.49009  &  2.03977      &  8   &  15.5029  &  8.06532   \\
3  & 7.15804  &  3.02782      &  9   &  17.1704  &  9.08144   \\
4  & 8.82745  &  4.02598      &  10  &  18.8418  &  10.0941   \\
5  & 10.4969  &  5.03039      &    &    &     \\
 \hline \hline
  \end{tabular}
\end{table}

\section{summary}

The wave equations for the Dirac fields in the SAdS black hole
spacetime are obtained by means of the Newman-Penrose formulism.
Then, the quasinormal frequencies corresponding to the Dirac field
perturbation in the SAdS black hole spacetime are evaluated by
using Horowitz-Hubeny approach and the results are presented by
tables and figures. We learn from the tables and figures that: (i)
For large black holes, both the real and the imaginary parts of
the fundamental quasinormal frequencies are linear functions of
the Hawking temperature, and the lines are described by $
Re(\omega)=8.367 T$, and $-Im(\omega)=6.371 T$, respectively.  For
intermediate and small size black holes, $Re(\omega)$ of the
fundamental quasinormal frequencies approximates the temperature
$T$ more closely than the black hole size $r_+$, and $Im(\omega)$
continue to lie along a straight line $-Im(\omega)=1.499 r_+ $.
(ii) For large black holes (say $r_+=100$), the fundamental
quasinormal frequencies depend very weakly on the low angular
quantum number (say $l=3/2$ and $l=5/2$). However, for
intermediate and small black holes, the angular quantum number $l$
mode has the surprising effect of increasing the damping time
scale (decreasing $|Im(\omega)|$) and decreasing the oscillation
time scale (increasing $Re(\omega)$). (iii) For large black holes,
an intermediate black hole ($r_+=1$) and a small black hole
($r_+=0.4$), the behave of the QNMs can be expressed as
$\frac{\omega(n)}{r_+}\sim (1.35-2.25 i) n +(2.136-1.500 i)$,
$\frac{\omega(n)}{r_+}\sim (2.02-2.32 i) n +(2.86-1.26 i)$ and
$\frac{\omega(n)}{r_+}\sim (4.18-2.48 i) n +(5.45-0.20 i)$,
respectively, which show that all the modes are evenly spaced, and
the smaller the black hole, the larger the spacing.

\begin{acknowledgments}This work was supported by the
National Natural Science Foundation of China under Grant No.
10275024 and under Grant No. 10473004; the FANEDD under Grant No.
200317; and the SRFDP under Grant No. 20040542003.
\end{acknowledgments}


\begin{thebibliography}{99}

\bibitem{Chand75} S. Chandrasekhar and S. Detweiler,
Proc. R. Soc. Lond.  {\bf A 344}, 441 (1975).

\bibitem{Hod} S. Hod,
Phys. Rev. Lett.  {\bf 81}, 4293 (1998).

\bibitem{Dreyer} O. Dreyer,
Phys. Rev. Lett.  {\bf 90}, 081301 (2003).


\bibitem{Baez} J. Baez, in Matters of gravity.   ed. J. Pullin, p.
12(Springer, 2003), gr-qc/0303027.

\bibitem{Kunstatter} G. Kunstatter, gr-qc/0212014; L. Motl, gr-qc/0212096;
A. Corichi, gr-qc/0212126; L. Motl and A. Neitzke,
hep-th/03301173; A. Maassen
 van den Brink, gr-qc/0303095.


\bibitem{Maldacena} J. Maldacena, Adv. Theor. Math. Phys.
  {\bf  2},  231 (1998).

\bibitem{Witten} E. Witten,  Adv. Theor. Math. Phys.
  {\bf  2},  253 (1998).

\bibitem{Kalyana} S. Kalyana Rama and Sathiapalan, Mod Phys. Lett.
 A {\bf  14},  2635 (1999).


\bibitem{Chan} J. S. F. Chan and R. B. Mann, Phys. Rev. D {\bf 55}  7546
(1997).

\bibitem{Birmingham}D. Birmingham, I. Sachs, and S. N. Solodukhin, Phys.
Rev. Lett. {\bf 88}  151301 (2002).

\bibitem{Cardoso} V. Cardoso and J. P. S. Lemos, Phys. Rev. D {\bf 63}
124015 (2001).

\bibitem{Konoplya} R. A. Konoplya, Phys. Rev. D {\bf 66}  084007 (2002).

\bibitem{Starinets} A. O. Starinets, Phys. Rev. D {\bf 66}  124013 (2002).

\bibitem{Horowitz} G. T. Horowitz and V. E. Hubeny, Phys. Rev. D {\bf 62}
024027 (2000).

\bibitem{Kurita} Y. Kurita and M. A. Sakagami, Phys. Rev. D {\bf 67}
024003 (2003).


\bibitem{Cardoso1} V. Cardoso and J. P. S. Lemos, Phys. Rev. D {\bf 64}
 084017 (2001).


\bibitem{Cardoso2} V. Cardoso, R. Konoplya, and J. P. S. Lemos,
gr-qc/0305037.

\bibitem{Konoplya1} R. A. Konoplya, Phys. Rev. D {\bf 66}  044009 (2002).


\bibitem{Page} D. N. Page, Phys. Rev. D 14, 1509  (1976).

\bibitem{Newman} E. Newman and R. Penrose, J. Math. Phys. (N. Y.) 3, 566
(1962).

\bibitem{Leaver} E. W. Leaver, Phys. Rev. D 34, 384 (1986).

\bibitem{Kokkotas} K. D. Kokkotas, Class. Quantum Grav. {\bf 8},
2217  (1991).

\bibitem{Cho} H. T. Cho, Phys. Rev. D 68, 024003 (2003).

\bibitem{Simone} L. Simone and C. M. Will, class. Quantum Grav.
{\bf 9}, 963  (1992).


\bibitem{Jing1} Jiliang Jing,   Phys. Rev. D 69, 084009 (2004).

\bibitem{Jing2} Jiliang Jing,  Phys. Rev. D 70, 065004 (2004).

\end{thebibliography}
\end{document}